\documentclass[aps,prl,twocolumn,showpacs,preprintnumbers,
               amsmath,amssymb]{revtex4}
\usepackage{graphicx}
\def\Comment#1{}
\newcommand{\bean}{\begin{eqnarray*}}
\newcommand{\eean}{\end{eqnarray*}}

\newcommand{\gapproxeq}{\lower
.7ex\hbox{$\;\stackrel{\textstyle >}{\sim}\;$}}
\newcommand{\lapproxeq}{\lower
.7ex\hbox{$\;\stackrel{\textstyle <}{\sim}\;$}}

\newcommand\lsim{\mathrel{\rlap{\lower4pt\hbox{\hskip1pt$\sim$}}
    \raise1pt\hbox{$<$}}}
\newcommand\gsim{\mathrel{\rlap{\lower4pt\hbox{\hskip1pt$\sim$}}
    \raise1pt\hbox{$>$}}}
\newcommand{\ba}{\begin{array}}
\newcommand{\ea}{\end{array}}
\newcommand{\nn}{\nonumber}

\newcommand{\be}{\begin{equation}}
\newcommand{\ee}{\end{equation}}
\newcommand{\bear}{\begin{eqnarray}}
\newcommand{\eear}{\end{eqnarray}}

\newcommand{\ket}{\,\rangle}
\newcommand{\bra}{\langle \,}

\newcommand{\cO}{{\cal O}}

\newcommand{\mL}{\mathcal{L}}

\newcommand{\Frac}[2]{\frac{\displaystyle #1}{\displaystyle #2}}

\begin{document}


\title{
Renormalizable Sectors in Resonance Chiral Theory:
\\
$S\to \pi\pi$ Decay Amplitude
 }

\author{L.Y. Xiao and J.J. Sanz-Cillero
}

\affiliation{
Department of Physics, Peking
University, Beijing 100871, P.R. China}

\email{xiaoly@pku.edu.cn,
cillero@th.phy.pku.edu.cn}

\date{\today}

\begin{abstract}
We develop a resonance chiral theory without any {\it a priori}
limitation on the number of derivatives in the hadronic operators.
Through an exhaustive analysis of the resonance lagrangian
and by means of field redefinitions, we find
that the number of independent operator contributing to the $S\to\pi\pi$ decay
amplitude is finite: there is only one single-trace operator (the
$c_d$ term) and three multi-trace terms. The deep implication of
this fact is that the ultraviolet divergences that appear in this
amplitude at the loop level can only appear through these chiral
invariant structures. Hence, a renormalization of these couplings
renders the amplitude finite.
\end{abstract}
\vskip .5cm


\maketitle


\section*{Introduction}

\vspace*{-0.35cm}
One of the most important issues in the construction
of hadronic lagrangians is how to restrict the number of operators
that contribute to a given
matrix element. The underlying symmetries of Quantum Chromodynamics (QCD)
constrain the structure of the action~\cite{lagrangian}
but the number of allowed symmetry invariant
terms is, in general, infinite.

At low energies, it is possible to describe the interaction of the Goldstones from
the spontaneous chiral symmetry breaking through an effective field theory based on
chiral symmetry, namely, Chiral Perturbation Theory ($\chi$PT)~\cite{chpt,U3-chpt,ChPT-op6}.
Although one may construct  an infinite number of operators, it is possible to
organise the lagrangian in terms of increasing number of derivatives, where
the dominant contribution to the low-energy amplitudes is provided by the lowest--order
operators.

As the energy is increased, heavier degrees of freedom (the mesonic resonances)
need to be included and the expansion in powers of momenta breaks down.
All the resonance operators are equally important,
independently of their number of derivatives.
Heuristically, it is possible to argue that terms with higher derivatives
spoil the asymptotic short-distance behaviour of
the QCD matrix elements~\cite{rcht,spin1fields}.
Although it has provided successful
phenomenological determinations~\cite{rcht,spin1fields,PI:02,VFFrcht,SSPPrcht},
this ``lowest-derivative-number'' rule still lacks of solid theoretical foundations
and, eventually, terms with higher derivatives may also need to be
considered~\cite{rcht-op6}.
Nonetheless, the crucial point to develop a resonance field theory is to prove
that for the description of any amplitude there is always a finite number of
independent operators
that enter into play.
A deep implication that stems from the existence of such minimal basis of operators
is that the structure of chiral-invariant ultraviolet divergences arising
at the loop level~\cite{generating}
must be also contained in this basis and, therefore, the number of counter-terms
for a definite amplitude is finite.

In this note, we analyse the general structure
of a chiral invariant theory for resonances where no {\it a priori} restriction
is made on the number of derivatives or resonance fields in the operators.
The large number of colours limit~\cite{NC1,NC3}
provides a convenient perturbative expansion in powers of $1/N_C$,
where loops appear at subleading orders~\cite{VFFrcht,SSPPrcht,generating}.
We focus the attention on the
terms of the lagrangian that contribute to the $S\to \pi\pi$ decay amplitude
and we find the corresponding minimal basis of operators by means of field redefinitions.

We want to stress again that the departing lagrangian
is completely general and no assumption
is made on the number of derivatives in the terms of the lagrangian.
No simplification is made on the part of the action responsible of other amplitudes,
which remains general all along the letter.
If a more complicate matrix element is to be computed, one should provide
again theoretical arguments to forbid higher derivative operators, independently
of whether the present $S\to\pi\pi$ simplifications are taken into account or not.

\vspace*{-0.5cm}
\section*{Building blocks of a chiral invariant action}

\vspace*{-0.35cm}
A non-linear realization of the
chiral Goldstone bosons is considered, being described
in the coset space by the coordinates
$\pi=\sum_a \frac{1}{\sqrt{2}}\lambda_a \pi_a $.
We choose the canonical coset representatives
$(\xi_L(\pi),\xi_R(\pi))$
such that $\xi_R(\pi)=\xi_L^\dagger(\pi)\equiv u(\pi)$~\cite{ChPT-op6}.
The latter transforms under the chiral
group $G$ in the way,
\begin{equation}
u(\pi) \,\,   \stackrel{g \in G}{ \longrightarrow } \,\,
g_{_R}\, u\, h^\dagger \,\, =\,\, h \, u\, g_{_L}^\dagger\,\,  ,
\end{equation}
with the exponential realization $u=\exp{\{ i \pi/\sqrt{2}\, F\} }$
and the chiral transformation $g=(g_{_L},g_{_R})\in G$.
The compensator field $h=h(g,\pi)$ depends both on the chiral transformation
and the Goldstone fields~\cite{ChPT-op6}.

The building blocks of our resonance lagrangian will be hadronic tensors
transforming covariantly under chiral transformations:
\vspace*{-0.45cm}
\begin{equation}
\label{eq.transform}
X\quad \stackrel{g\in G}{ \longrightarrow} \quad h\, X\, h^\dagger\,.
\end{equation}
We choose a representation where the
$q \overline{q} $ resonance multiplets $S,V...$ transform in this way~\cite{rcht,rcht-op6}.
The Goldstone fields will enter in the lagrangian through
the basic covariant tensors~\cite{ChPT-op6},
\begin{eqnarray}
 u_\mu &=&  i \, \{ u^\dagger (\partial_\mu - i r_\mu) u  \,
-\, u \, (\partial_\mu- i \ell_\mu) u^\dagger \}\, ,
\nn
\\
\chi_\pm &=& u^\dagger \, \chi\, u^\dagger \, \pm \, u\, \chi^\dagger \, u \, ,
\label{eq.bricks}
\\
 f_\pm^{\mu\nu} &=& u\, F_L^{\mu\nu}\, u^\dagger \, \pm \, u^\dagger \, F_R^{\mu\nu}\, u\, ,
 \nn
\end{eqnarray}
or covariant derivatives  $\nabla^\rho\nabla^\sigma...$ of them.
The field $\chi= 2B_0(s+i p)$ contains the scalar and pseudo-scalar
external sources, $s$ and  $p$ respectively.
The $f_\pm^{\mu\nu}$ are proportional to $r^\mu$ and $\ell^\mu$ sources, which
provide the vector and axial-vector external sources, $v^\mu=\frac{1}{2}(r^\mu+\ell^\mu)$
and $a^\mu=\frac{1}{2}(r^\mu -\ell^\mu)$ respectively~\cite{chpt,U3-chpt,ChPT-op6}.

The covariant derivative is given by~\cite{ChPT-op6,rcht,rcht-op6}
\begin{equation}
\nabla_\mu\, X\, =\, \partial_\mu \, X\, \, +\, \, [\Gamma_\mu\, , \, X]\, ,
\end{equation}
with the chiral connection
$\Gamma_\mu=\frac{1}{2}\{ u^\dagger\, (\partial_\mu - i r_\mu) u
\, +\, u\, (\partial_\mu- i \ell_\mu) u^\dagger \}$.
The commutation properties of the covariant derivatives
will be used in the next section~\cite{ChPT-op6}:
\begin{eqnarray}
[ \nabla^\mu ,\nabla^\nu ]  X &=& [ \Gamma_{\mu\nu} , X ]\, ,
\\
&&\quad \mbox{with}\quad \Gamma_{\mu\nu}\, =\,
\Frac{1}{4} [u_\mu,u_\nu] \, - \, \Frac{i}{2} f_{+\, \mu\nu}\, .
\nn
\end{eqnarray}
Every time  the order of two covariant derivatives is exchanged
we generate an extra operator given by $\Gamma_{\mu\nu}$,  which is proportional
to either $v_\mu$ and $a_\mu$ ($f_{+\, \mu\nu}$ term)
or to at least two $u^\alpha$ tensors ($[u_\mu,u_\nu]$ term), i.e.,
$ \Gamma_{\mu\nu}\sim \cO(J) + \cO(u^\mu u^\nu)$.
All along the letter, we will denote any term proportional to at least
one source $J=s,\, p,\, v^\mu,\, a^\mu$  as $\cO(J)$.

\vspace*{-0.5cm}
\section*{Resonance chiral theory lagrangian}

\vspace*{-0.35cm}
We will call Resonance Chiral Theory (R$\chi$T) to the most general
chiral invariant hadronic action. No {\it a priori} restriction is made on the
number of derivatives in the lagrangian operators.
Its  basic building blocks $X$ are the resonance fields $R=S,V...$,
the Goldstone tensors $u^\mu$, $\chi_\pm$, $f_\pm^{\mu\nu}$, and covariant derivatives
$\nabla^{\alpha_1}...\nabla^{\alpha_n} X$ of any of them.
A priori, symmetry does not impose any constraint on the number of derivatives
or resonance fields in the operators of the lagrangian. It only determines the
way how the hadronic fields must be combined~\cite{lagrangian,rcht,rcht-op6}.

Nevertheless, we provide in the next lines two important simplifications that relie
on the freedom to redefine the hadronic fields in the generating
functional $W[J]$~\cite{rcht-op6}.

\vspace*{0.15cm}
{\bf a.) Goldstone field redefinitions}

Any R$\chi$T lagrangian can be expressed in the general form,
\vspace*{-0.cm}
\begin{eqnarray}
\mL &=& \Frac{F_0^2}{4}\bra u_\mu u^\mu\ket \,
+\, \bra A_S\, \nabla^\mu u_\mu\ket \,
+\, \bra B_S\ket
\nn
\\
&&\quad
\, -\, \Frac{1}{2}\bra S (\nabla^2 +M_S^2) S\ket
\, +\, \Delta \mL
\, ,
\label{eq.La}
\end{eqnarray}
with $\bra ...\ket$   short for trace in flavour space
and where the remaining part of the lagrangian contains the terms that will
not enter in our problem:
\begin{eqnarray}
\Delta\mL &=&  \cO(S^2 u^\alpha u^\beta) \, + \cO(S^{3})
\nn
\\
&&\,\,\,  +\, \cO( R')
\, + \, \cO(J)\, +\,\cO(u^\alpha u^\beta u^\mu u^\nu)\,  .
\label{eq.DLa}
\end{eqnarray}
The term $\cO(S^{3})$ refers to operators with at least three scalar fields and
$\cO(R')$ to terms containing at least one resonance
$R'\neq S$. At least one external source is contained in the operators $\cO(J)$.
At leading order in $1/N_C$ (LO),
the only  operator bilinear in the scalar fields is the canonical kinetic term
and those with two scalars must be either $\cO(S^2 u^\alpha u^\beta)$ or $\cO(J)$.
Only at next-to-leading order (NLO),
other $S$--bilinear operators will be allowed, although their presence
is not relevant for the argumentation on the Goldstone field redefinition.
Notice that the last two terms in Eq.~(\ref{eq.DLa})
may contain both resonances and Goldstone fields, accounting for
the remaining $\chi$PT--like operators  allowed by the symmetry.

The tensors $A_S,\, B_S$ in Eq.~(\ref{eq.La}) are hermitic and linear in the scalar field.
For later convenience, we define them such that they cannot be included in any of the
last three terms in Eq.~(\ref{eq.DLa}), i.e.,
$\bra A_S \nabla^\mu u_\mu\ket ,\, \bra B_S\ket
\neq \cO(R')+ \cO(J)+ \cO(u^\alpha u^\beta u^\mu u^\nu)$.
We gather in the term $\bra B_S\ket$ the  linear operators in $S$ of this kind
that cannot be written like  $\bra A_S\nabla^\mu u_\mu\ket$. From this definitions,
we have that both $\bra A_S \nabla^\mu u_\mu\ket$ and $\bra B_S\ket$
can just include one scalar field and two tensors
$u^\mu$ --or any number of covariant derivatives of them--, i.e.,
$A_S\sim S u^\alpha$ and $B_S\sim S u^\alpha u^\beta$. Their explicit form will
be provided in the next section.

We will perform at this point a Goldstone field redefinition that induces
a shift in $u^\mu$ of the form
\begin{equation}
u_\mu \,\,\longrightarrow \,\, u_\mu \,\, +\,\, \Frac{2}{F_0^2}\, \nabla_\mu A_S
\,\, +\,\, \cO(A_S^2)\, .
\end{equation}
The required Goldstone transformation is not unique. One could consider, for instance,
${   \xi_R\to \xi_R \exp{\{ - i A_S/F_0^2\} }  ,    }  \,\,
   {   \xi_L\to \xi_L \exp{\{ i A_S/F_0^2 \} }     }$.
Notice that $A_S/F_0^2$ is indeed dimensionless and hermitic  and
the $\xi_{R,L}$ remain unitary.

This transformation produces a lagrangian with exactly
the same structure in Eqs.~(\ref{eq.La})--(\ref{eq.DLa})
except for  the term $\bra A_S\nabla^\mu u_\mu\ket$, which is completely removed.

\vspace*{0.15cm}
{\bf b.) Scalar field redefinitions}

After the transformation in section a), one gets the simplified lagrangian,
\begin{eqnarray}
\mL &=& - \Frac{1}{2}\bra S (\nabla^2+M_S^2) S\ket
\,+\, \bra S (\nabla^2 +M_S^2) \,\zeta \ket \, +\, \bra S \,\eta \ket
\nn
\\
&& \qquad \, +\, \Frac{F_0^2}{4}\bra u^\mu \, u_\mu\ket \, +\, \Delta \mL
\, ,
\label{eq.Lb}
\end{eqnarray}
where we have made the replacement
\begin{eqnarray}
\bra B_S\ket \,\, &=&\,\,\bra S (\nabla^2 +M_S^2) \, \zeta \ket \, +\, \bra S \, \eta \ket \, .
\end{eqnarray}
The tensor $\bra S \, \eta\ket$ gathers all the terms in $\bra B_S\ket $
which cannot be expressed like $\bra S (\nabla^2+M_S^2)\, \zeta\ket$, and
the structure of $\Delta \mL$ was defined in Eq.~(\ref{eq.DLa}).

Now we perform the  convenient field redefinition,
\vspace*{-0.1cm}
\begin{equation}
S\quad \longrightarrow \quad S \,+ \,   \, \zeta\, ,
\end{equation}
which yields  the  simplified lagrangian
\vspace*{-0.1cm}
\begin{eqnarray}
\mL' &=& - \Frac{1}{2}\bra S (\nabla^2+M_S^2) S\ket \, +\, \bra S \eta \ket
\\
&&
\quad  + \, \Frac{F_0^2}{4}\bra u^\mu \, u_\mu\ket
\, +\,  \Delta\mL \, ,
\nn
\end{eqnarray}
where the term $\bra S (\nabla^2 +M_S^2) \, \zeta \ket$ has been fully removed from the action.
We have made use of the fact that $\eta,\, \zeta\sim \cO(u^\alpha u^\beta)$.

A final detail is that beyond LO, in addition to the kinetic term, one  could consider
a subleading operator in the lagrangian
of the form $\lambda  \bra S(\nabla^2+M_S^2)^k S\ket $, with $k\geq 2$. The shift
$S\to S + \lambda (\nabla^2+M_S^2)^{k-1} S$ removes this term, leaving  remainders
of this same form at higher subleading orders. Hence, by iteration,
we can always move the operators of this form to higher orders in perturbation theory, beyond
any the order we are working at. Thus, the decomposition of the lagrangian
given in Eqs.~(\ref{eq.La}),~(\ref{eq.DLa})~and~(\ref{eq.Lb}) is indeed general.

\vspace*{-0.5cm}
\section*{General form of the $S\pi\pi$ chiral operators}

\vspace*{-0.35cm}
In the chiral limit,
all the operators that contribute to the $S\to \pi\pi$ decay amplitude
are contained in the terms $\bra A_S \nabla^\mu u_\mu\ket +\bra B_S\ket$
of  the initial lagrangian in Eq.~(\ref{eq.La}).
However, we have seen that the operators of the form
 $\bra A_S \nabla^\mu u_\mu\ket$ and $\bra S (\nabla^2+M_S^2) \, \zeta \ket$
are not physical and they
can be always removed through a convenient change in
the meson field variables.

We will write now the explicit form of these operators.
They cannot contain the tensors $\chi_\pm$ and $f_\pm^{\mu\nu}$ since
these are proportional to external sources. Hence, they can be only composed of
the tensors $S$ and $u^\mu$, or any number of  derivatives of them.
In the absence of external sources, one has that $u^\mu$ is proportional to at least one
Goldstone field so our operator may contain at most two tensors of this kind.
Finally, through partial integration
we can always move the derivatives away from the scalar field. Hence, a general term
contributing to $S\to \pi\pi$ takes the form
\vspace*{-0.1cm}
\begin{eqnarray}
\mL_{S\to\pi\pi}
\, &=& \, \lambda \, \bra S\, \left\{\, \nabla^{\mu_1}\, ...\, \nabla^{\mu_m}\, u^\rho\, ,
\, \nabla^{\nu_1}\, ...\, \nabla^{\nu_n}\, u^\sigma \right\}\ket
\nn
\\
&& \qquad \,\,\times
\, \, t_{\mu_1, ...\, \mu_m, \, \rho, \, \nu_1,...\, \nu_n,\, \sigma}\quad ,
 \label{eq.general-Spipi}
\end{eqnarray}
where the Lorentz tensor $t_{\mu_1, ...\, \mu_m, \, \rho, \, \nu_1,...\, \nu_n,\, \sigma}  $
handles all the contractions of the indices. The anticommutator $\{...\, ,\, ...\}$
ensures that the operator is invariant under charge and hermitian
conjugations~\cite{ChPT-op6}. The number of covariant derivatives $m,\, n$
can be any positive integer number or  zero.
Any reordering of the covariant derivatives $\nabla^{\mu_i}\nabla^{\mu_{i+1}}$ in
$\nabla^{\mu_1}...\nabla^{\mu_m}$  (or similarly for
$\nabla^{\nu_j}\nabla^{\nu_{j+1}}$ in
$\nabla^{\nu_1}...\nabla^{\nu_n}$) generates an extra operator containing
the chiral tensor $\Gamma^{\mu_i\mu_{i+1}}$, which does not contribute to
$S\to \pi\pi$ since it contains $f_+^{\mu_i\mu_{i+1}}$ or a number
of $u^\alpha$ tensors higher than two. Hence, we will freely change
the order of the covariant derivatives for the convenience of the derivation.

The simplest operator of this kind is
\begin{eqnarray}
\mL_{S\to \pi\pi}\, \, =\, \, \lambda \, \bra \, S\, \{ u^\mu \, , u_\mu \}\, \ket
\,\, =\,\,  2\, \lambda \bra S\,  u^\mu \, u_\mu\ket \, ,
\label{eq.cd}
\end{eqnarray}
which is just the $c_d \bra S u^\mu u_\mu \ket$ operator in Ref.~\cite{rcht}.

For a higher number of derivatives,
the available contractions of the Lorentz indices yield four possible  cases, where
one index $\mu_i$ can be contracted with $\rho$, with another $\mu_j$, with some index $\nu_j$
or with $\sigma$:

{\bf 1.) For $m\geq 1$, there can be
at least one of the indices $\mu_i$ contracted with $\rho$:}
The covariant derivative $\nabla^{\mu_i}$ can be commuted until it is placed
next to $u^\rho$ in Eq.~(\ref{eq.general-Spipi}). Hence, this case is
equivalent to contracting the last index $\mu_m$ and $\rho$:
\begin{eqnarray}
&&\mL_{S\to\pi\pi}\, \, =\, \,
\nn
\\
&&\qquad \lambda
\bra S\, \left\{\, \nabla^{\mu_1}\, ...\, \nabla^{\mu_{m-1}}\,
( \nabla^{\rho}u_\rho) \, ,
\, \nabla^{\nu_1}\, ...\, \nabla^{\nu_n}\, u^\sigma \right\}\ket
\nn
\\
&& \qquad \,\,\times
\, \, t_{\mu_1, ...\, \mu_{m-1}, \, \nu_1,...\, \nu_n,\, \sigma}\quad .
\end{eqnarray}
Due to the $\nabla^\rho u_\rho$ tensor in the operator,
this term does not contribute to the $S\to \pi\pi$ amplitude
in the chiral limit when the out-going Goldstones are on-shell. Furthermore,
as we saw in the previous section, this  operator can be fully removed from the lagrangian
through  an appropriate  Goldstone field redefinition.

This case is equivalent to that when  $n\geq 1$
and at least one of the indices $\nu_j$ is contracted with $\sigma$.

{\bf 2.)  For  $m\geq 2$, there can be
at least one of the indices $\mu_i$ contracted with another index $\mu_j$:}
The covariant derivatives $\nabla^{\mu_i}$ and  $\nabla^{\mu_j}$ can be commuted
until both of them are placed next to $u^\rho$. Hence, this case is
equivalent to contracting the last index $\mu_{m-1}$ and $\mu_{m}$:
\begin{eqnarray}
&& \mL_{S\to\pi\pi}
\, \, =
\nn
\\
&&\qquad  \, \lambda\,
\bra S\, \left\{\, \nabla^{\mu_1}\, ...\, \nabla^{\mu_{m-2}}\,
\nabla^2 u^\rho \, ,
\, \nabla^{\nu_1}\, ...\, \nabla^{\nu_n}\, u^\sigma \right\}\ket
\nn
\\
&& \qquad \,\,\times
\, \, t_{\mu_1, ...\, \mu_{m-2}, \, \nu_1,...\, \nu_n,\, \sigma}\quad ,
\end{eqnarray}
This can be converted into the former case by means of the identity
\begin{eqnarray}
\nabla^2 u^\rho &=& \nabla^\rho \, (\nabla^\alpha u_\alpha) - \nabla_{\alpha} f_-^{\alpha\rho}
+ [\Gamma^{\alpha\rho},u_\alpha]
\nn
\\
&&  = \nabla^\rho \, (\nabla^\alpha u_\alpha)\,  +\,  \cO(J)\, +\,
\cO(u^\alpha u_\alpha u^\rho)\, ,
\end{eqnarray}
and, therefore, it does not contribute $S\to \pi\pi$.

This case is equivalent to that when  $n\geq 2$
and at least one of the indices $\nu_i$ is contracted with another~$\nu_j$.

{\bf 3.)  For $m,n \geq 1$, there can be
at least one of the indices $\mu_i$ contracted with one of the
$\nu_j$:}
We can commute the covariant derivatives $\nabla^{\mu_i}$ and  $\nabla^{\nu_j}$ and
move them to the front part of the operator. Hence, this case is
equivalent to contracting the first indices, $\mu_1$ and $\nu_1$:
\begin{eqnarray}
&& \mL_{S\to\pi\pi} \,\, =
\nn
\\
&&\qquad   \lambda
\bra S\, \left\{\, \nabla^{\mu}\,\nabla^{\mu_2}\, ...\, \nabla^{\mu_m}\, u^\rho\, ,
\, \nabla_{\mu}\,  \nabla^{\nu_2}\, ...\, \nabla^{\nu_n}\, u^\sigma \right\}\ket
\nn
\\
&& \qquad \,\,\times
\, \, t_{\mu_2, ...\, \mu_m, \, \rho, \, \nu_2,...\, \nu_n,\, \sigma}\quad .
\label{eq.case3}
\end{eqnarray}
Through partial integration, it can be converted into the expression
\begin{eqnarray}
&& \mL_{S\to\pi\pi}\,\, =
\nn
\\
&&\qquad   \Frac{\lambda}{2}\left[
\bra \nabla^2 S\, \left\{\,\nabla^{\mu_2}\, ...\, \nabla^{\mu_m}\, u^\rho\, ,
\,   \nabla^{\nu_2}\, ...\, \nabla^{\nu_n}\, u^\sigma \right\}\ket \right.
\nn
\\
&&\qquad \quad
-\, \bra  S\, \left\{\, \nabla^2\,\nabla^{\mu_2}\, ...\, \nabla^{\mu_m}\, u^\rho\, ,
\,  \nabla^{\nu_2}\, ...\, \nabla^{\nu_n}\, u^\sigma \right\}\ket
\nn
\\
&& \qquad \quad \left.
-\, \bra  S\, \left\{\, \nabla^{\mu_2}\, ...\, \nabla^{\mu_m}\, u^\rho\, ,
\, \nabla^2\,   \nabla^{\nu_2}\, ...\, \nabla^{\nu_n}\, u^\sigma \right\}\ket  \right]
\nn
\\
&& \qquad \,\,\times
\, \, t_{\mu_2, ...\, \mu_m, \, \rho, \, \nu_2,...\, \nu_n,\, \sigma}\quad .
\end{eqnarray}
The second and third terms reproduce the case no.~2 and
do not contribute to $S\to \pi\pi$.
The first term can be rewritten as
\begin{eqnarray}
&&\mL_{S\to\pi\pi}  \,\, =
\nn
\\
&& \,\,\Frac{\lambda}{2} \left[
\,  - \,
M_S^2\, \bra  S\, \left\{\,\nabla^{\mu_2}\, ...\, \nabla^{\mu_m}\, u^\rho\, ,
\,   \nabla^{\nu_2}\, ...\, \nabla^{\nu_n}\, u^\sigma \right\}\ket
\right.
\nn
\\
&& \left.+
\bra S (\nabla^2+M_S^2)\, \left\{\,\nabla^{\mu_2} ... \nabla^{\mu_m}\, u^\rho\, ,
  \nabla^{\nu_2} ... \nabla^{\nu_n}\, u^\sigma \right\}\ket
\right]
\nn
\\
&& \qquad \,\,\times
\, \, t_{\mu_2, ...\, \mu_m, \, \rho, \, \nu_2,...\, \nu_n,\, \sigma}\quad .
\label{eq.case3B}
\end{eqnarray}
The last term can be removed through
the scalar field redefinition in previous section,
so the only non-vanishing contribution to $S\to\pi\pi$
comes from  the first term in Eq.~(\ref{eq.case3B}).
This leave us with an operator that shows again the functional structure
in Eq.~(\ref{eq.general-Spipi}), and where
the number of derivatives has been  decreased in two orders.
Hence, it admits to be reanalysed
and eventually further simplified.

{\bf 4.)   In the last case remaining,  for $m,n \geq 1$, there can be
one of the indices $\mu_i$ contracted with $\sigma$ and  one of the
$\nu_j$ contracted with $\rho$:}
We can commute the covariant derivatives  $\nabla^{\mu_i}$ and  $\nabla^{\nu_j}$  and
move them backwards  until they are placed next to $u^\rho$ and $u^\sigma$,
respectively. Hence, this case is
equivalent to contracting the last indices,
$\mu_{m}$ with $\sigma$  and $\nu_n$ with $\rho$:
\begin{eqnarray}
&&\mL_{S\to\pi\pi}\,\, =
\nn
\\
&&\qquad  \lambda
\bra S\, \left\{\, \nabla^{\mu_1}\, ...\, \nabla^{\mu_{m-1}}
\, \nabla^{\mu} u^\nu\, ,
\, \nabla^{\nu_1}\, ...\, \nabla^{\nu_n}\,  \nabla_{\nu} u_\mu \right\}\ket
\nn
\\
&& \qquad \,\,\times
\, \, t_{\mu_1, ...\, \mu_{m-1}, \, \nu_1,...\, \nu_{n-1}}\quad .
\end{eqnarray}
One can apply the chiral tensor relation~\cite{ChPT-op6},
\begin{equation}
\nabla_\nu u_\mu = \nabla_\mu u_\nu + f_{-\, \mu\nu}\, =\,
\nabla_\mu u_\nu +\cO(J)\, ,
\end{equation}
where the $\cO(J)$  term does not contribute to $S\to\pi\pi$.
Finally, moving the $\nabla^\mu$
covariant derivatives to the front of the operator, we get
the structure analysed in the case no.~3 in Eq.~(\ref{eq.case3}).

This completes the list of operators that may contribute
to the $S\to \pi\pi$ decay amplitude. All non-vanishing terms can be written
in the way shown in the case no.~3 and then simplified into an operator
with a lower number of derivatives. By iteration, it is then possible to convert any
term into the $c_d$ operator in Eq.~(\ref{eq.cd}),
up to contributions irrelevant for $S\to \pi\pi$.

In fact, if one admits $1/N_C$--suppressed  operators in the reasoning, there are
three more available --multitrace-- terms:
$\bra S \ket \bra u^\mu u_\mu\ket$, $\bra S u^\mu\ket \, \bra u_\mu\ket$ and
$\bra S\ket\, \bra u^\mu\ket\, \bra u_\mu\ket$. This finally exhausts the list of
$S\pi\pi$ operators, both at LO in $1/N_C$ and at subleading orders.

\section*{Conclusions}

\vspace*{-0.35cm}
The example in this note provides a first clear example of the possibility of
constructing model independent resonance chiral lagrangians. Although the action may contain
an infinite number of operators, the particular vertex functions could be described
through a finite number of them, making the theory predictable and model independent.
In this article,
the $S\to\pi\pi$ decay amplitude is determined by one single-trace operator in the
chiral limit (four if we include the subleading multi-trace terms).
The remaining part of the action is kept fully general both before and after simplifications.

A deeper implication of the present result refers to the structure of the
loop ultraviolet divergences in the generating functional $W[J]$~\cite{generating}.
In the case of the  $S\to\pi\pi$ vertex function, they    can  only
take the form of these four chiral invariant local operators.
Hence, this amplitude can be rendered finite through a renormalization of
$c_d$ and the other three subleading couplings.
The full theory may require an infinite number of coupling renormalizations
but only a finite set of them is required for the study of particular amplitudes.

\vspace*{-0.5cm}
\section*{Acknowledgments}

\vspace*{-0.35cm}
We want to acknowledge useful comments and discussions with
K.~Kampf, B.~Moussallam, J.~Novotny, J.~Portol\'es, P.~Ruiz-Femen\'\i a, M.X.~Su and
H.Q.~Zheng.
  This work is supported in part by National
 Nature Science Foundations of China under contract number
 10575002,
  10421503.

\vspace*{-0.6cm}


\begin{thebibliography}{90}









\bibitem{lagrangian}
  C.G. Callan {\it et al.},
  Phys. Rev. {\bf 177} (1969) 2247; \\
%
%
  S.R. Coleman {\it et al.},
  Phys. Rev. {\bf 177} (1969) 2239;




\bibitem{chpt}
    J. Gasser and H. Leutwyler,
    Ann. Phys.~{\bf 158} (1984) 142;
 \\
    Nucl. Phys. B {\bf 250} (1985) 465.




\bibitem{U3-chpt}
  R.~Kaiser \& H.~Leutwyler,
  Eur. Phys. J. C {\bf 17} (2000) 623;
  \\
%
%
  P.~Herrera-Siklody {\it et al.},
  Nucl. Phys. B {\bf 497} (1997) 345.




\bibitem{ChPT-op6}
    J. Bijnens {\it et al.},
    JHEP {\bf 9902} (1999) 020.




\bibitem{rcht}
  G. Ecker {\it et al.},
  Nucl. Phys. B {\bf 321} (1989) 311.


\bibitem{spin1fields}
  G.~Ecker {\it et al.},
  Phys. Lett. B {\bf 223} (1989) 425.




\bibitem{PI:02}
    A. Pich [arXiv:hep-ph/0205030].





\bibitem{VFFrcht}
    I. Rosell {\it et al.},
    JHEP  {\bf 0408} (2004) 042.



\bibitem{SSPPrcht}
    I. Rosell {\it et al.},
    JHEP  {\bf 0701} (2007) 039;







\bibitem{rcht-op6}
  V.~Cirigliano {\it et al.},
  Nucl. Phys. B {\bf 753} (2006) 139.

\bibitem{generating}
    I. Rosell {\it et al.},
    JHEP  {\bf 0512} (2005) 020.


\bibitem{NC1}
  G. 't Hooft, Nucl. Phys. B {\bf 72} (1974) 461.


\bibitem{NC3}
  E. Witten, Nucl. Phys. B {\bf 160} (1979) 57.
























\end{thebibliography}
\end{document}